\documentclass[aps,prl,twocolumn,showpacs,floatfix]{revtex4}

\usepackage{latexsym}
\usepackage{graphicx}
\newcommand{\la}{\langle}
\newcommand{\ra}{\rangle}
\newcommand{\nicefrac}[2]{\leavevmode\kern.1em
    \raise.5ex\hbox{\the\scriptfont0 #1}\kern-.1em
    /\kern-.15em\lower.25ex\hbox{\the\scriptfont0 #2}}
\begin{document}
\title{Using the
    Wigner-Ibach Surmise to Analyze Terrace-Width Distributions:
    History, User's Guide, and Advances}
\author{T. L. Einstein}
\email[Email: ]{einstein@umd.edu}
\homepage[Homepage: ]{http://www2.physics.umd.edu/~einstein/}
\affiliation{Department of Physics, University of Maryland, College Park, MD 20742-4111}
\begin{abstract}
A history is given of the applications of the simple expression generalized from the surmise by Wigner and also by Ibach to extract the strength of the interaction between steps on a vicinal surface, via the terrace width distribution (TWD).  A concise guide for use with experiments and a summary of some recent extensions are provided.
\end{abstract}
\pacs{68.35.Md, 05.40.-a, 68.37.Ef}
\maketitle

{\it Dedicated to Prof. Harald Ibach, with profound gratitude, on the occasion of his retirement}
\vspace{-5mm}
\section{Introduction}

Measurement of the terrace width distribution (TWD) $\check{P}(\ell)$ of vicinal
surfaces is now used routinely to find the dimensionless strength
$\tilde{A}$ of the elastic repulsion between steps.  The University of Maryland \cite{JW99} and the Forschungszentrum-J\"ulich \cite{Giesen01} have been at the vanguard of this progress.  Use of an extension of the Wigner surmise from random matrix theory has resolved ambiguities on how best to estimate $\tilde{A}$ from the variance.  This paper discusses the history of this development and the crucial role played by Harald Ibach in divining from physical insight an analytic expression for the TWD associated with the special case $\tilde{A}\! = \! 0$ that is essentially the same as the simple expression that Wigner surmised describes (albeit, ultimately, not exactly) the distribution of the energy differences of adjacent levels in a nucleus when the coupling has unitary symmetry.  This insight spawned my long collaborative effort with FZ-J\"ulich to develop generalizations appropriate to arbitrary vicinal surfaces and to corroborate the viability of the resulting 
formulation to account for extensive experimental data.  In addition to a personal history of this progress (with a few new results), I collect (with enhancements) in one place several tables from publications and present a concise ``User's Guide" for applying the formalism and a short discussion of some recent developments, e.g., for azimuthally misoriented vicinal surfaces and for non-equilibrium situations.  Space limitations preclude fuller discussions or an authoritative update on experimental progress.  Three of the four figures are heretofore unpublished.  Also interwoven are comments stimulated by several penetrating questions, many by Harald Ibach, that arose during talks I gave in Germany that forced me to sharpen my thinking.  These are topics typically skirted in publications as being well-established, but my experience is that they are not generally well understood.  The use of ``we" below is not a polite affectation but instead reflects the crucial role played by various of my many collaborators in this extended series of studies.

\begin{figure}[t]
\includegraphics[width= 8.5cm,viewport=80 50 690 510,clip]{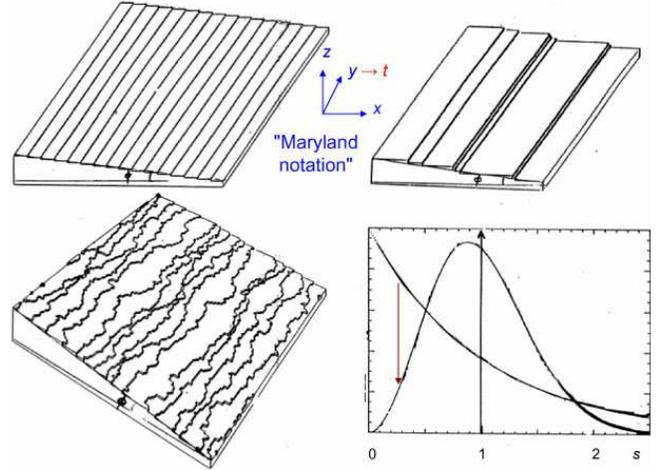}
\caption {Plots of generic lattice configurations and associated TWDs, to illustrate essential features, as discussed in text, of steps on a vicinal surface (with no energetic interactions).  Top left: ``perfect" cleaved crystal.  Top right: straight steps placed randomly, corresponding to a strictly 1D model.  Lower left: meandering steps in a TSK model.  Lower right: TWDs $P(s)$ associated with these distributions.  The downward arrow emphasizes the greatly decreased chance of finding steps at very small separation when they meander, as compared to straight steps, due to the entropic repulsion.
}
{\label{f:hiTWD}}
\end{figure}

To set the stage (cf.\ Fig.~\ref{f:hiTWD}), the direction perpendicular to the terraces (which are densely-packed facets) is typically called $\hat{z}$.  In ``Maryland notation" the normal to the vicinal surface lies in the $x-z$ plane, and the distance $\ell$ between steps is measured along $\hat{x}$, while the steps run along the $\hat{y}$ direction.  In the simplest and usual approximation, the repulsions between adjacent steps arise from two sources: an entropic or steric interaction due to the physical condition that the steps cannot cross, since overhangs cannot occur in nature.  The second comes from elastic dipole moments due to local atomic relaxation around each step, leading to frustrated lateral relaxation of atoms on the terrace plane between two steps.  Both interactions are proportional to $1/\ell^2$.  

To illustrate the essence of $\check{P}(\ell)$, we consider in Fig.~\ref{f:hiTWD} its shape for 3 idealized configurations.  There is just one characteristic length in the $\hat{x}$ direction, namely the average step separation $\la \ell \ra$.  (Contrary to widespread misconception, $\la \ell \ra$ need not be a multiple of, or even simply related to, the substrate lattice spacing. It is the step height times $\cot \phi$, where $\phi$ is the {\it arbitrary} misorientation, as in shown in Fig.~\ref{f:hiTWD}.)  Therefore, we consider $P(s)= \la \ell \ra^{-1}\check{P}(\ell)$, where $s\equiv \ell/\la \ell \ra$, a dimensionless length.  For a ``perfect" cleaved crystal, $P(s)$ is just a spike $\delta (s-1)$.  If, as intrinsic to 1D models, the steps are imagined as uncooked spaghetti dropped at any position with probability $1/\la \ell \ra$, $P(s)$ is a Poisson distribution $\exp(-s)$.  Actual steps do meander, as one study most simply in a terrace-step-kink (TSK) model.  In this model, the only excitations are kinks (with energy $\epsilon$) along the step.  (This is a good approximation at low temperature $T$ since adatoms or vacancies on the terrace cost several $\epsilon$ [4$\epsilon$ in the case of a simple cubic lattice].  The entropic repulsion due to step meandering dramatically decreases the probability of finding adjacent steps at $\ell \ll \la \ell \ra$.  To preserve the mean of one, 
$P(s)$ must also be smaller than $\exp(-s)$ for large $s$.

If there is an additional energetic repulsion $A/\ell^2$, the magnitude of the step meandering will decrease, narrowing $P(s)$.  As $A \rightarrow \infty$, the width approaches a delta function.  Note that the energetic and entropic interactions do not simply add.  In particular, there is no negative (attractive) value of $A$ at which the two cancel each other.  Thus, for strong repulsions, steps rarely come close, so the entropic interaction plays a smaller role, while for $A<0$, the entropic contribution increases, as illustrated in Fig.~\ref{f:enteng} and worked out in detail below.

\section{History}

Investigation of the interaction between steps on vicinal surfaces is a core part of the flourishing field of exploring the properties of these technologically important and scientifically rich systems, as discussed in several excellent reviews \cite{JW99,Giesen01,Noz}.
The earliest studies seeking to extract $A$ from TWDs used the mean-field-like Gruber-Mullins \cite{Gruber67} approximation, in which a single active step fluctuates between two fixed straight steps $2\la \ell \ra$ apart.  Then the energy associating with the fluctuations $x(y,t)$ is 
\begin{equation}
 \Delta {\cal E} = -\beta(0)L_y + \int_0^{L_y} \beta(\theta(y))\sqrt{1+\left(\frac{\partial x}{\partial y}\right)^2} dy ,
\label{e:fluc}
\end{equation}
\noindent where $\beta$ is the step free energy per length (or line tension \cite{ibachstress}) for a step at orientation $\theta$ relative to the mean direction of the step (and the direction of the fixed, bounding steps), and $L_y$ is the size of the system along the mean step direction (i.e. the step length with no kinks).  We expand $\beta(\theta)$ as the Taylor series $\beta(0) + \beta^\prime(0)\theta + 
\nicefrac{1}{2}\,\beta^{\prime\prime}(0)\theta^2$ and recognize that the length of the line segment has increased from $dy$ to $dy/\cos \theta \approx dy(1 + \nicefrac{1}{2}\,\theta^2)$.  For close-packed steps, for which $\beta^\prime(0)=0$, it is well known that  (using $\theta \approx \tan \theta = \partial x/\partial y$),  

\begin{figure}[t]
\includegraphics[bb= 87 0 375 183, width=8.5cm]{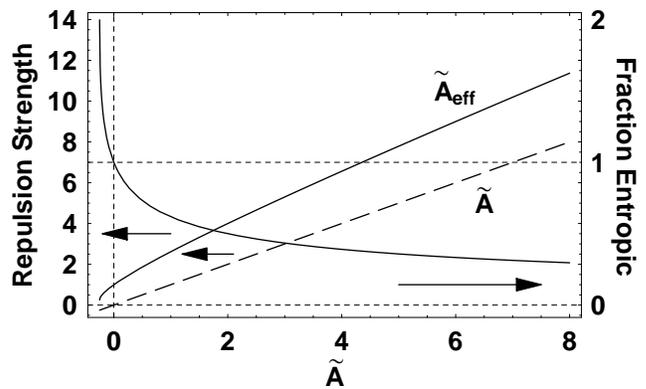}
\caption {Illustration of how entropic repulsion and energetic interactions combine, plotted vs.\ the dimensionless energetic interaction strength $\tilde A \equiv a\tilde \beta/(k_BT)^2$.  The dashed straight line is just $\tilde A$.  The solid curve above it is the combined entropic and energetic interactions, labeled $\tilde A_{\rm eff}$ for reasons explained below.  The difference between the two curves at any value of the abscissa is the dimensionless entropic repulsion for that $\tilde A$. The decreasing curve, scaled on the right ordinate, is the ratio of this entropic repulsion to the total dimensionless repulsion $\tilde A_{\rm eff}$.  It falls monotonically with $\tilde A$, passing through unity at $\tilde A = 0$.  See the discussion accompanying Eq.~(\ref{e:Aeff}) for more information and explicit expressions for the curves. 
}
{\label{f:enteng}}
\end{figure}

\begin{equation}
 \Delta {\cal E} \approx \frac{\tilde\beta(0)}{2}\int_0^{L_y}\! \left(\frac{\partial x}{\partial y}\right)^2 \! dy, \quad  \tilde\beta(0) \equiv \beta(0)+\beta^{\prime\prime}(0),
\label{e:stiff}
\end{equation}
\noindent where $\tilde\beta$ is the step stiffness \cite{FFW}. Most treatments gloss over the fact that the stiffness has the same definition for steps with arbitrary in-plane orientation.  The key point is that to create such steps, one must apply a ``torque" \cite{Leamy} which exactly cancels the linear term $\beta^\prime(0)\theta$.  Stasevich \cite{timThesis} provides a more formal proof.

Since $x(y)$ is taken to be a single-valued function that is defined over the whole domain of $y$, the 2D configuration of the step can be viewed as the worldline of a particle in 1D by recognizing $y$ as a time-like variable.  Since the steps cannot cross, these particles can be described as 1D fermions.  We also see from Eq.~(\ref{e:stiff}) that in the (1+1)D formulation, $\partial x/\partial y \rightarrow \partial x/\partial t$, with the form of a velocity, so that the stiffness plays the role of a {\it mass}; indeed, it serves as the inertial parameter of steps in this fermion perspective (though not with regard to actual dynamics in response to external forces \cite{AP-FP}). Moreover, stiffness is one of the three ingredients of the very-successful step-continuum model \cite{JW99}.  

Pursuing this analogy for polymers in 2D, de Gennes \cite{deG} showed nearly 4 decades ago that this problem could be mapped into the Schr\"odinger equation in 1D, with the thermal energy $k_BT$ replacing $\hbar$.  Then in the Gruber-Mullins approximation \cite{Gruber67}, the step with no energetic interactions becomes a particle in a 1D infinite well of width $2\la \ell \ra$, with well-known groundstate properties

\begin{equation}
 \psi_0(\ell) \!  \propto \sin\left(\frac{\pi \ell}{2\la \ell \ra}\right); \; P(s)=\sin^2\left(\frac{\pi s}{2}\right); \; E_0= \frac{(\pi k_BT)^2}{8\tilde\beta\la \ell \ra^2}
\label{e:1Dfree}
\end{equation}
\noindent Thus, it is the kinetic energy of the ground state in the quantum model that corresponds to the entropic repulsion (per length) of the step.  In the exact solution for the free energy expansion of the equilibrium crystal shape \cite{AAY}, the numerical coefficient in the corresponding term is $\nicefrac{1}{6}$ rather than $\nicefrac{1}{8}$.  Note that $P(s)$ peaks at $s=1$ and vanishes for $s\ge 2$.

Suppose, next, that there is an energetic repulsion $U(\ell) = A/\ell^2$ between steps.  In the 1D Schr\"odinger equation, the prefactor of $-\partial^2 \psi(\ell)/\partial \ell^2$ is $(k_BT)^2/2\tilde\beta$; hence, $A$ only enters the problem in the dimensionless combination $\tilde{A} \equiv A\tilde\beta/(k_BT)^2$ \cite{JWks99}.  In the Gruber-Mullins picture, the potential (per length) experienced by the single active particle is (with $\check\ell \equiv \ell - \la \ell \ra$):
\begin{equation}
\tilde{U}(\check\ell) = \frac{\tilde{A}}{(\check\ell\! -\! \la \ell \ra)^2} + \frac{\tilde{A}}{(\check\ell \! +\! \la \ell \ra)^2}
= \frac{2\tilde{A}}{\la \ell \ra^2} +\frac{6\tilde{A}\check\ell^2}{\la \ell \ra^4} +\frac{10\tilde{A}\check\ell^4}{\la \ell \ra^6} +\ldots
\label{e:sho}
\end{equation}

\noindent The first term is just a constant shift in the energy.  For $\tilde{A}$ sufficiently large, the particle is confined to a region $|\check\ell| \ll \la \ell \ra$, so that we can neglect the fixed walls and the quartic term, reducing the problem to the familiar simple harmonic oscillator, with the solution:

\begin{equation}
\! \psi_0(\ell) \!  \propto {\rm e}^{-\check\ell^2/4 w^2}; \; 
 P_{\rm G}(s) \equiv \frac{1}{\sigma_G \sqrt{2\pi}}
              \exp\left[-\frac{(s-1)^2}{2\sigma_G^2}\right]
\label{e:1Dsho}
\end{equation}
\noindent where $\sigma_G = (48\tilde A)^{-1/4}$ and $w =\sigma_G \la \ell \ra$.

For $\tilde A$ of 0 or 2, it turns out, as we shall see, that the TWD can be computed exactly.  For these cases, Eqs.~(\ref{e:1Dfree}) and (\ref{e:1Dsho}), respectively, provide serviceable approximations.  It is Eq.~(\ref{e:1Dsho}) that is prescribed for analyzing TWDs in the most-cited resource on vicinal surfaces \cite{JW99}.  Indeed, it formed the basis of initial successful analyses of experimental STM data \cite{WW}.  However, it has some notable shortcomings.  Perhaps most obviously, it is useless for small but not vanishing $\tilde A$, for which the TWD is highly skewed, not resembling a Gaussian, and the peak, correspondingly, is significantly below the mean spacing.  For large values of $\tilde A$, it significantly underestimates the variance or, equivalently, the value of $\tilde A$ one extracts from the experimental TWD width:  Ihle et al. \cite{IMP98} point out that in the Gruber-Mullins approximation the TWD variance is the same as that of the active step, since the neighboring step is straight.  For large $\tilde A$ the fluctuations of the individual steps on an actual vicinal surface become relatively independent, so the variance of the TWD is the {\it sum} of the variance of each, i.e. twice the step variance.  Given the great (quartic) sensitivity of $\tilde A$ to the TWD width, this is problematic.  As experimentalists acquired more high-quality TWD data, other approximation schemes were proposed, all producing Gaussian distributions with widths $\propto \tilde A^{-1/4}$, but with proportionality constants notably larger than $48^{-1/4}= 0.38$. 

For the ``free-fermion" ($\tilde A = 0$) case, Jo\'os et al. \cite{Joos91} developed at the beginning of the 1990's a sequence of analytic approximants to the exact but formidable expression \cite{d62,MehtaRanMat} for $P(s)$.  The procedure is based on a discrete version of the problem, representing the fermion-like steps in terms of second-quantized operators and taking note that the TWD is not just a pair-wise step-step correlation function but actually a many-particle correlation function in which we demand that no step lie between the steps at 0 and $\ell$.  The $n$th approximant behaves well at smaller $s$ but eventually passes through 0 around $s=n$ and then behaved non-physically.  Specifically, the leading behavior of each approximant is $1 - [\sin(\pi s)/(\pi s)]^2 \approx \nicefrac{1}{3}(\pi s)^2$.
In the asymptotic region, a different approach shows leading behavior for large $s$ is $P(s) \propto s^{7/4}\exp[-\nicefrac{1}{8}(\pi s)^2]$, where the proportionality constant has the numerical value $\sim$9.46.  (The analytic form is in Ref.~\cite{Joos91}.)   This expression works to surprisingly small $s$, reproducing the peak semi-quantitatively.  (It would have gone to even smaller $s$ than indicated in Fig.~4 of Ref.~\cite{Joos91} had the two leading asymptotic corrections not been included!)  Finally, Ref.~\cite{Joos91} as well as a slightly earlier paper \cite{Bartelt90} draw the analogy between the TWD of vicinal surfaces and the distribution of spacings between interacting (spinless) fermions on a ring, the Calogero-Sutherland model \cite{C69,Sutherland71}, 
which in turn for three particular values of the interaction---in one case repulsive ($\tilde{A} = 2$), in another attractive ($\tilde{A} = -\nicefrac{1}{4}$), and lastly the free-fermion case ($\tilde{A} = 0$)---could, astonishingly, be solved exactly by connecting to random matrix theory \cite{d70,MehtaRanMat};   Fig.~5 of Ref.~\cite{Joos91} depicts the three resulting TWDs.  

\begin{figure}[t]
\scalebox{0.055}{\includegraphics{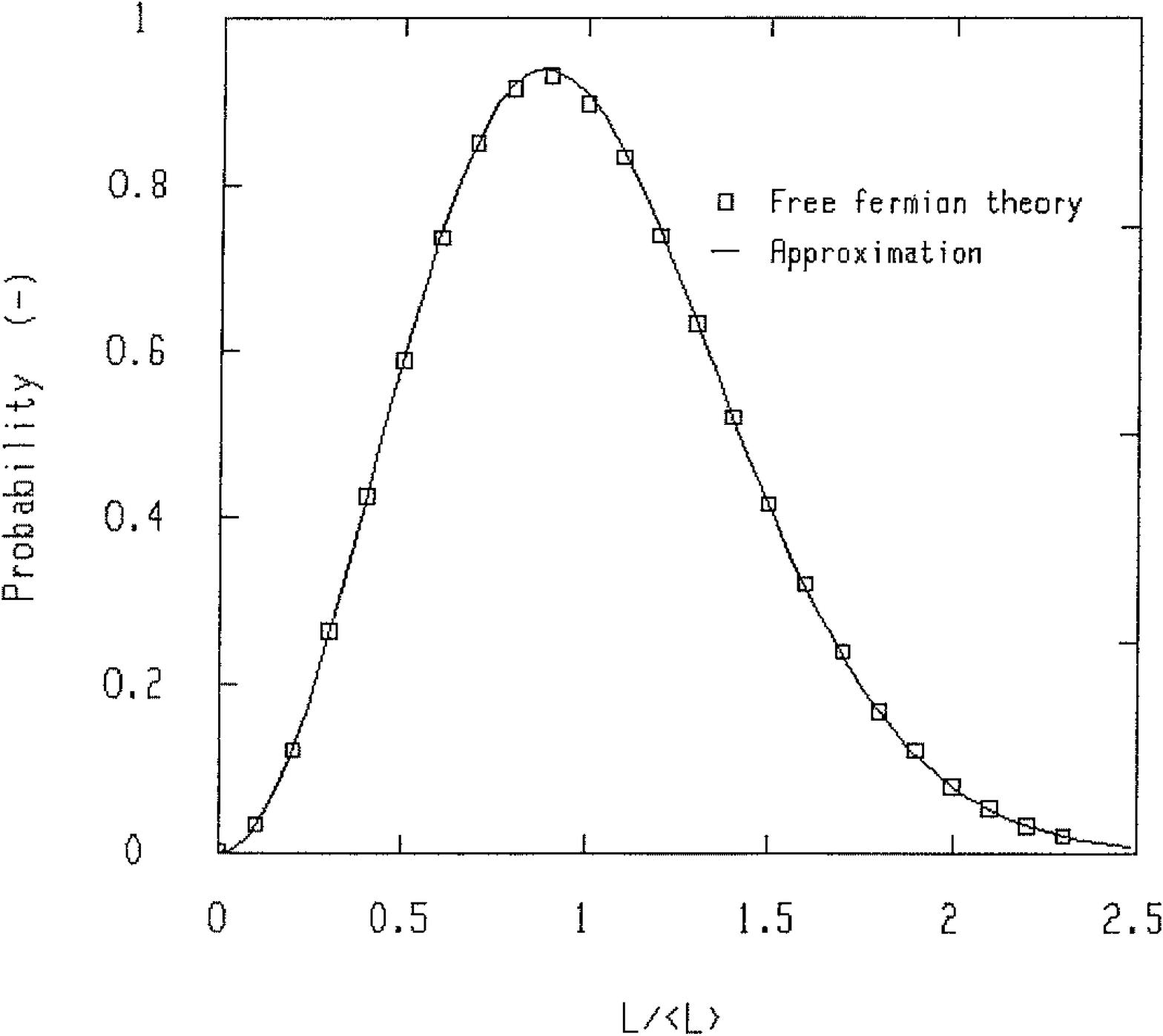}}
\caption {The graph for the expression deduced by H. Ibach for the TWD of steps with no energetic repulsion (solid curve), essentially Eq.~(\ref{e:w2}) surmised by Wigner, while the points are generated using the formalism for the approximants, given in Ref.~\cite{Joos91}
}
{\label{f:WI}}
\end{figure}

This was the state of affairs when Harald Ibach presented me (seemingly sometime during his sabbatical stay in College Park during 1996--7) with Fig.~\ref{f:WI}, in which he achieved outstanding agreement with [numerical approximation visually indistinguishable from] the exact solution by plotting what essentially amounts to 
\begin{equation}
P_2(s)=\frac{32}{\pi^2} s^2 \exp(-\frac{4}{\pi} s^2),
\label{e:w2}
\end{equation}
 
\noindent where the subscript of $P$ refers to the exponent of $s$, and the pair of numerical factors ensure that the distribution is normalized and has a mean of one.  He challenged me to explain his insight, and Fig.~\ref{f:WI} graced my office wall, greeting me each morning, for well over a year.  Not hearing back from me, he put the remarkable expression (albeit with some typos that obscured its potency) into Ref. \cite{Giesen97}.  On Feb. 2, 1998, Safi Bahcall gave an intriguing condensed-matter seminar at University of Maryland on ``Superconductivity and Random Matrices" (cf.~Ref.~\cite{safi}). In subsequent discussions he rifled through a preprint of Guhr {\it et al.}'s seminal review of random matrix theory \cite{Guhr98}, and I spotted Fig.~12 therein, noting the similarity to Fig.~5 in Ref.~\cite{Joos91}, as well as the accompanying discussion of the Wigner surmise, their Eq.~(3.50) being the first I had seen of what will be Eq.~(\ref{e:Wigner}) below.

The other two curves correspond to corresponding Wigner surmises for those cases:

\begin{equation}
 P_4(s) = \left(\frac{64}{9\pi}\right)^3 s^4 \exp\left( -\frac{64}{9\pi}s^2\right);
\label{e:w4}
\end{equation}
\begin{equation} 
P_1(s) = \frac{\pi}{2} s^1 \exp\left( -\frac{\pi}{4}s^2\right).
\label{e:w1}
\end{equation}
\noindent In random matrix literature, the exponent of $s$, viz.\ 1, 2, or 4, is called $\beta$, due to an analogy with inverse temperature in one justification.  To avoid possible confusion with the step free energy per length $\beta$ or the stiffness $\tilde\beta$ for vicinal surfaces, we have called it instead by the Greek symbol that looked most similar, $\varrho$.  The Appendix offers transparent arguments on how the three kinds of symmetry lead to the associate exponents 1, 2, and 4. 
As seen most clearly by explicit plots, e.g.\ Fig.~4.2a of
Haake's text \cite{Haake91}, $P_1(s)$, $P_2(s)$, and $P_4(s)$ are
excellent approximations of the exact results for orthogonal,
unitary, and symplectic ensembles, respectively, and these simple
expressions are routinely used when confronting experimental data in a broad range of physical problems
\cite{Guhr98,Haake91}.  (The agreement is particularly outstanding
for $P_2(s)$ and $P_4(s)$, which are the germane cases for vicinal
surfaces: the variance is 1\% below and 0.4\% above the exact values, respectively; for $P_1(s)$ it is 4-1/2\% below.  This agreement is significantly better than any other approximation (cf.\ Table~2 of Ref.~\cite{Hailu}) and far better than the Gruber-Mullins approximation, as depicted in Fig.~1 of Ref.~\cite{EP99}, Fig.~1 of Ref.~\cite{ERCP01}, and Fig.~2 of Ref.~\cite{E03}.)

\begin{table*}[t]
\caption{Tabulation of various measurable properties of terrace-width
distributions $P(s)$ [where $s$ is the terrace width normalized by its
average value] based on exact results at the three soluble values of the
dimensionless interaction strength
$\tilde{A}$, the corresponding generalized Wigner distribution (GWD) expression, and the various Gaussian approximations: Gruber-Mullins (GM), modified Grenoble \cite{IMP98,EP99,PM98} (G), and Saclay \cite{Barbier96}(S). (In the original Grenoble approximation \cite{IMP98,PM98}, $\sigma^2 \propto \tilde{A}^{-1/2}$ rather than $\tilde{A}_{\rm eff}^{-1/2} 
\equiv 2/\varrho$, but with the same prefactor as indicated in this table.)  
SHO $\Rightarrow$ simple harmonic oscillator,
i.e.\ uniformly spaced steps [energies].  The extreme case
$\varrho$=0, for which exact results are trivial, is included to
dramatize trends in $\varrho$ \cite{l}.  As
$\varrho$ increases, the TWD becomes narrower, more symmetric, and more nearly
Gaussian.  Anticorrelations of neighboring terrace-width fluctuations increase.  For
the three exactly-solvable [non-trivial] cases, the GWD provides an
excellent approximation, far better than any alternative.  [Adapted from Ref.~\cite{EP99}, with most results for the ``Exact" case from Refs.~\cite{Guhr98,MehtaRanMat}, with new evaluations for symplectic case from Ref.~\cite{HE06}.]
}
\vspace{0.1cm}
\begin{tabular}{cc|ccccc} 
 Property & Case & $\varrho =0$ & $\varrho
=1$ &
$\varrho =2$ &
$\varrho =4$ &$\varrho \rightarrow \infty$\\
    &  & ``Random" & Attractive & Non-interact & Repulsive & Extreme rpl. \\
\multicolumn{2}{c|} {\raisebox{1.1ex}[0pt]
{\fbox{\fbox{$\tilde{A} = \frac{\varrho}{2}(\frac{\varrho}{2}-1)$}$  \quad \tilde{A}_{\rm eff} = \left( \frac{\varrho}{2}\right)^2 = \tilde A +\frac{\varrho}{2}$}}} &
$\tilde{A} = 0^-$&$\tilde{A} = - 1/4$ & $\tilde{A} =0$ & $\tilde{A}
=2$&$\tilde{A} = \varrho^2/4$
\\  \hline
 \multicolumn{2}{c|} {Underlying ${\cal H}$ symmetry} &[Poisson]
& orthogonal & unitary  & symplectic &  [SHO+phonons] \\
\hline
Variance& Exact & 1 & 0.286 & 0.180 & 0.1041 &$0^+$\\ 
$\mathbf{\sigma^2} =\mu_2$&GWD: $[(\varrho + 1)/2b_\varrho] -1$  & 0.5708 & 0.2732 &0.1781 & 0.1045 &
$0.500 \varrho^{-1}$\\
 \hspace{0.2cm} $=\mu^{\prime}_2 -1$& GM (all): $\sqrt{15/8\pi^4} \doteq 0.139 \tilde{A}^{-1/2}$ & --- & --- &
0.1307 & 0.0981 &
$0.278
\varrho^{-1}$\\ 
 & GM (NN): $1/\sqrt{48} \doteq 0.144 \tilde{A}^{-1/2}$ & --- & --- & 0.1307 & 0.1021 & $0.289
\varrho^{-1}$\\
 \hline 
Alternative & G (all): $\! \! \frac{1}{\pi}\! \int_0^{2\pi} \! \! \! \! \frac{1-\cos{\phi}}{\phi(2\pi-\phi)}\!  \doteq\!  0.247 \tilde{A}_{\rm eff}^{-1/2}$& --- & --- & --- & 0.1747 & $0.495
\varrho^{-1}$\\
 estimate & G (NN): $\! \sqrt{2/3\pi^2}\!  \doteq\!  0.260  \tilde{A}_{\rm eff}^{-1/2}$& --- & --- & --- & 0.1838 & $0.520
\varrho^{-1}$\\
of $\sigma^2$ & S: $2/\pi^2 \doteq 0.203 \tilde{A}_{\rm eff}^{-1/2}$   & --- & --- & 0.203 & 0.101 &
$ 0.405 \varrho^{-1}$\\
\hline
Neighboring&Exact cov$^a$ ($ s_1,s_2$) & 0& $-0.27$ & $-0.31$ & $-0.34$& \\
 terraces&Exact $\langle (s_1 + s_2 -2)^2\rangle$ & 2 & 0.416 & 0.248 &
0.138 &$0^+$\\ \hline
Peak&Exact &0 & 0.77 & 0.8840 & 0.941 & $1^-$\\ 
position&GWD: $(\varrho/2b_\varrho)^{1/2}$ & 0& 0.7979 & 0.8862
 & 0.9400  &$1 - 0.250 \varrho^{-1}$\\
 $s_{\rm max}$ &Gruber-Mullins & --- & --- & 1 & 1 &1\\ \hline 
Skewness &Exact  & 2 &  & 0.4972 & 0.350(1) &\\ 
 $\mu_3/\sigma^3=$&GWD: $\mu_3^\prime=(\varrho + 2)/2b_\varrho$   & 0.9953 & 0.6311 & 0.4857 & 0.3542 & $0.707
\varrho^{-1/2}$\\
 $(\mu_3^{\prime} \!- \! 1)/\sigma^{3} \!-
\!3/\sigma$& Gaussian Approx.   & --- & --- & 0 & 0& 0\\
\hline Kurtosis&Exact &9 & & 3.1 & 3.027(1) & \\ 
$\mu_4/\sigma^4$ &GWD: $\mu_4^\prime/\sigma^4 =(\varrho+3)/(\varrho+1)$ & 3.8691 &3.2451 & 3.1082 & 3.0370 & $3+
0.750\varrho^{-2}$\\
 &Gruber-Mullins & ---  & --- & 2.4062 & 3 & 3\\ \hline  
\end{tabular}
$^a$Covariance of adjacent terrace widths: cov$(s_1,s_2)\equiv\frac{\langle (s_1-\langle
s_1\rangle)(s_2-\langle s_2\rangle )\rangle} {\left[\langle (s_1-\langle
s_1\rangle)^2\rangle \langle (s_2-\langle s_2\rangle )^2\rangle
\right]^{1/2}} =\sigma^{-2}[\langle s_1s_2\rangle -1]= -1+\langle (s_1 + s_2
-2)^2\rangle/2\sigma^2$
\end{table*}

With the values $\varrho$ = 1, 2, and 4. the three specific expressions Eqs.~(\ref{e:w2}--\ref{e:w1}) comprising the so-called Wigner
surmise \cite{Guhr98,Haake91} can be written as a single formula
\begin{equation}
  \label{e:Wigner}
   P_\varrho(s) =
    a_\varrho s^{\varrho} \exp \left(-b_\varrho s^2\right) \, ,
\end{equation}
where the constants
$b_\varrho$, which fixes its mean at
unity, and $a_\varrho$, which normalizes $P(s)$, are
\begin{equation}
    b_\varrho =
  \left[\frac{\Gamma \left(\frac{\varrho +2}{2}\right)}
             {\Gamma \left(\frac{\varrho +1}{2}\right)}\right]^2  
\quad
    a_\varrho = \frac{2\left[\Gamma
\left(\frac{\varrho +2}{2}\right)\right]^{\varrho +1} }
                  { \left[\Gamma
\left(\frac{\varrho +1}{2}\right)\right]^{\varrho +2} } =
\frac{2b_\varrho^{(\varrho+1)/2}}{\Gamma \left(\frac{\varrho +1}{2}\right)} \, .
\label{e:abr}
\end{equation}

As noted above the Calogero-Sutherland model provides a connection between random matrix results, notably the Wigner surmise, and the distribution of spacings between fermions in 1D interacting with dimensionless strength $\tilde A$.  Specifically, 
\begin{equation}
  \label{e:avrh}
  \tilde{A} = \frac{\varrho}{2}\left(\frac{\varrho}{2} -1\right)
     \quad \Leftrightarrow \quad \varrho = 1 +\sqrt{1 +4 \tilde{A}}.
\end{equation}
For an arbitrary system, there is no reason that $\tilde{A}$ should take on one of the three special values.  However, it seems impossible to generalize the arguments in the Appendix to general values.  For our purposes, the obvious solution is to use Eq.~(\ref{e:avrh}) for arbitrary $\varrho$ or $\tilde{A}$.  Curiously, this form has not been applied in conventional investigations involving RMT, but instead other analytic phenomenological expressions, e.g.\ those proposed by Brody \cite{Brody}
and by Izrailev \cite{Izra}
are used (cf. Ref.~\cite{Guhr98}).  Such applications typically involves mixtures of systems of two of the symmetries; neither they nor other models involving crossover between well-defined symmetries \cite{SPBT} are analogous to systems based on the Calogero-Sutherland model for arbitrary $\varrho$. 

For arbitrary $\varrho$ there is no symmetry-based
justification of distribution based on the Wigner surmise of Eq.\ (\ref{e:Wigner}).  Nonetheless,
we have argued that it provides a viable, arguably optimal interpolation scheme
between the two special
values of $\varrho$ and also out to the Grenoble expression for nearly-infinite
repulsion;\cite{EP99,ERCP01} we have also used it successfully to analyze experimental
data.\cite{ERCP01,Giesen00,RCEG00}
For brevity, we refer hereafter to this set of formulas, Eqs.\
(\ref{e:Wigner},\ref{e:abr}) as the GWD (generalized Wigner surmise); elsewhere \cite{Hailu,ERCP01,RCEG00} we have called it CGWD (continuum generalized Wigner
distribution).

The moments of the GWD can be
expressed simply in terms of $b_\varrho$:

\begin{equation}
  \mu_n^\prime \equiv \int_0^\infty \! \! a_\varrho s^{\varrho+n} \exp \left(-b_\varrho s^2\right) ds
  =\frac{\Gamma \left(\frac{\varrho +1+n}{2}\right)}{b_\varrho^{n/2}\Gamma \left(\frac{\varrho +1}{2}\right)}
\label{e:momt}
\end{equation}
 
\noindent The variance $\sigma^2_W = \mu_2 = \mu_2^\prime -1$ of the GWD is then

\begin{equation}
   \sigma^2_W = \frac{\varrho + 1}{2b_\varrho} - 1
             \approx  \frac{1}{2\varrho}
                      - \frac{3}{8\varrho^2}
                      + \frac{3}{16\varrho^3}
                         - \frac{7}{384\varrho^4}
                         + {\cal O}(\varrho^{-5})
\label{e:sr}
\end{equation}

\noindent for large values of
$\varrho$, as given in Eq.\ (A8) of Ref.\ \cite{RCEG00}.  Note that $b_\varrho$ cancels in the  ratio: 
\begin{equation}
  \frac{\mu_n^\prime}{(\mu_2^\prime)^{n/2} }
  =\frac{\Gamma \left(\frac{\varrho +1+n}{2}\right)}{\left(\frac{\varrho+1}{2}\right)^{n/2}\Gamma \left(\frac{\varrho +1}{2}\right)}
\label{e:mrat}
\end{equation}

Similarly, the peak of $p_\varrho(s)$ (i.e.\ its mode) occurs at
\begin{equation}
   s_{\rm max} = \left( \frac{\varrho}{2b_\varrho}\right)^{1/2}
             \approx 1- \frac{{\rm e}^{-\varrho}}{4} -\frac{1}{4\varrho}
\label{e:sm}
\end{equation}
For the special values of $\varrho$, the peak positions are tabulated in Table 1.  We further note that $s_{\rm max}$ = 0.96, 0.97, and 0.98 occur for $\varrho$= 6.1, 8.2, and 12.4, respectively.  When dealing with experimental data, such shifts from $s_{\rm max}$ =1 would be difficult to distinguish. 

While there are some formal justifications for the GWD as an optimal description of TWDs for arbitrary $\tilde{A}$, arguably the most convincing argument is a comparison of the predicted variance with numerical data generated from Monte Carlo simulations.  We include in the comparisons some Gaussian approximations (viz.\ approximation schemes which lead to TWDs that are Gaussians), alluded to earlier, e.g.\ in Table 1.  For these the dimensional variance of the TWD must scale like $\la \ell \ra^2$.  The approximations differ in how the dimensionless part depends on $\tilde A$.  The approximation developed in  Grenoble by Ihle, Pierre-Louis, and Misbah \cite{IMP98,PM98} focuses on the limit of large $\varrho$, neglecting the entropic interaction in that limit.  While the variance $\sigma^2 \propto \tilde A^{-1/2}$, the proportionality constant is 1.8 times that in the Gruber-Mullins case (cf. Table 3).  One improve this approximation, especially for repulsions that are not extremely strong, by including the entropic interaction in an average way.  This is done by replacing $\tilde A$ by

\begin{equation}
  \label{e:Aeff}
  \tilde{A}_{\rm eff} = \left( \frac{\varrho}{2}\right)^2 = \tilde A +\frac{\varrho}{2}.
\end{equation}
\noindent (Cf.\ Eq.~(\ref{e:avrh}).)  In this modified scheme, $\sigma^2 \propto \varrho^{-1}$.

The physical meaning of $\tilde{A}_{\rm eff}$ has not been adequately discussed heretofore.  It corresponds to the full strength of the inverse-square repulsion between steps, i.e. the modification due to the inclusion of entropic interactions.  From Eq.~(\ref{e:Aeff}) it is obvious that the contribution of the entropic interaction, viz. the difference between the total and the energetic interaction, as discussed in conjunction with Fig.~(\ref{f:enteng}), is $\varrho/2$.  This quantity, as noted then, depends sensitively on $\tilde{A}$.  Note also that, remarkably, the ratio of the entropic interaction to the total interaction is $(\varrho/2)/(\varrho/2)^2=2/\varrho$; this is the fractional contribution that is plotted in Fig.~\ref{f:enteng}.

Barbier et al. at Saclay \cite{Barbier96,Masson94,LeGoff99} consider instead the $m$-terrace width relative to $m\la \ell \ra$, comparing the variance for large $m$ to the coefficient of $\ln m$ expected from roughening theory, they conclude that $\sigma^2 = 4/(\pi^2 \varrho)$, i.e.\ the same form as the modified Grenoble approximation but with a proportionality constant $K$ that is 82\% as large.  This information is summarized in Table 3.  Note that for the Saclay and Wigner approaches, one must assume all steps interact with an inverse square repulsion; for the others, one can treat either that assumption or allow only nearest neighbors to interact energetically.

\begin{figure}[t]
\includegraphics[bb= 110 5 315 145, width= 3.3in]{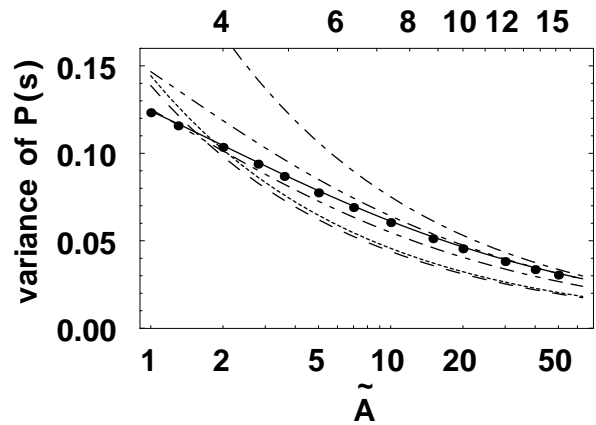}
\caption[shrt1]{
The variance $\sigma^2(\tilde{A})$ vs.\ $\tilde{A}$ (and vs.\ $\varrho$ on the upper abscissa) on a logarithmic scale,
plotted for the GWD (light solid curve) and for the various Gaussian approximations:
the modified Grenoble  (
short-short-long dashed for all steps interacting, short-long dashed for NN step interactions only), 
Saclay (short-short-long-long dashed line), and Gruber-Mullins 
(long dashed for all steps interacting, short dashed
for NN steps only).   Monte Carlo data are shown as $\bullet$'s, with statistical errors
less than the size of the symbols.  See text for discussion.  [Fig.~2a of Ref.~\cite{Hailu}]
}
\protect\label{lin-log}
\end{figure}

In Fig.~\ref{lin-log} we compare how well all these theoretical approximations account for the behavior extracted from a well-controlled numerical experiment, Metropolis Monte Carlo simulations \cite{NewBark} of a TSK model.  For this model (with $A$=0), the
characteristic distance between close approaches of neighboring steps is \cite{BEW92}
\begin{equation}
y_{\rm coll} \! = \! \langle \ell
\rangle^2 \tilde{\beta}/4k_BT \! = \! (\langle \ell \rangle /2)^2
\sinh^2(\epsilon/2k_BT),
  \label{e:ycoll}
\end{equation}
which is about 35 at $k_BT/\epsilon = 1/2$ and
$\langle \ell \rangle \! = \! 10$, used in our most extensive calculations. For $A\! > \!0$, meandering is suppressed, making this distance is larger.  In most cases, we used $L_y \! = \! 2000 \gg y_{\rm coll}$, and  $N \! = \! 100$ steps \cite{MC50}.
We used a standard high-quality random-number generator (Ran3
\cite{NR}) and averaged over 100 runs using different initial seeds.  In these runs
the variance reached its steady-state value after about 3000 MCS (Monte Carlo steps per site); we started
``taking data" after $10^4$ MCS, recording results every 10 MCS until reaching
$3\! \times \!10^4$ MCS \cite{Hailu}.

The excellent agreement between the GWD expression and numerical data generated with Monte Carlo simulations is displayed in Fig.~\ref{lin-log}.  The various
predictions of the variance are plotted vs.\ $\tilde{A}$.  A logarithmic scale is
used for the horizontal axis so as not to give undue visual emphasis to larger
values of
$\tilde{A}$ nor to blur the region of rapid variation for small (but non-vanishing)
$\tilde{A}$, for which an exact calibration point exists.  The Wigner result is
essentially given by Eq.\ (\ref{e:as}).  Table 3 shows that the physical values of $\tilde{A}$ range from near 0 up to the mid teens.  Some pathology is presumably involved in the rare
cases in which larger values are observed.  There are relatively few reports of small but
non-zero values of
$\tilde{A}$.  A reason might well be that any of the Gaussian
approximations manifestly fail in this regime, so that before the recognition of
the utility of the Wigner distribution, one could not deal quantitatively with
small $\tilde{A}$ \cite{SBV}.

\begin{table}
\begin{tabular}{lc|c|ccc}  
Vicinal & $T$(K) & $\sigma^2$ & $\varrho$ & $\tilde{A}$ & 
  $W \! /\! G$ \\
\hline
Pt(110)-(1$\times$2) \cite{SBV}& 298 & & 2.2 & 0.13 & --- \\ 
Cu (19,17,17) \cite{RCEG00,GS99}& 353 & 0.122 & 4.1 & 2.2 & 0.77\\ 
Si(111)  \cite{BMHF}& 1173 & 0.11 & 3.8 & 1.7 & 0.96\\
Cu(1,1,13)  \cite{RCEG00,Giesen97}& 348 & 0.091 & 4.8 & 3.0 & 1.27 \\
Cu(11,7,7) \cite{RCEG00,GS99} & 306 & 0.085 & 5.1 & 4 & 1.37 \\
Cu(111) \cite{RCEG00,GS99} & 313 & 0.084 & 5.0 & 3.6 & 1.39 \\
Cu(111) \cite{RCEG00,GS99} & 301 & 0.073 & 6.0 & 6.0 & 1.58 \\
Ag(100) & 300 & 0.073 & 6.4 & 6.9 & 1.58 \\
Cu(1,1,19) \cite{RCEG00,Giesen97}& 320 & 0.070 & 6.7 & 7.9 & 1.64  \\
Si(111)-(7$\times$7) \cite{W94} & 1100 & 0.068 & 6.4 & 7.0 & 1.67 \\ 
Si(111)-(1$\times$1)Br  \cite{WW}& 853 & 0.068 & 6.4 & 7.0 & 1.67 \\ 
Si(111)-Ga \cite{FKI} & 823 & 0.068 & 6.6 & 7.6 & 1.67\\ 
Si(111)-Al $\surd3$  \cite{SW}& 1040 & 0.058 & 7.6 & 10.5  & 1.85 \\ 
Cu(1,1,11) \cite{Barbier96}& 300 & 0.053 & 8.7 & 15  & 1.95 \\ 
Cu(1,1,13) \cite{RCEG00,Giesen97} & 285 & 0.044 & 10 & 20 & 2.12 \\  
Pt(111)  \cite{HSMFK}& 900 & 0.020 & 24 & 135 & 2.59\\ 
\hline  
\end{tabular}
\caption{Compendium of experiments measuring the variances of terrace width
distributions of vicinal systems, as of the end of the last decade.
The estimate of $\tilde{A}$
is obtained from the (normalized) variance using Eq.~(\ref{e:as}), except for the first-row entry,
which is based on a direct fit using 
the 2-parameter GWD (cf.\ 6th item of ``User's Guide"). $W/G$ stands for the ratio of the estimates of the interaction strength based on Wigner and Gruber-Mullins perspectives, $\tilde{A}_W/\tilde{A}_G$, as given in Eq.~(\ref{e:aa}). [Condensed from Ref.~\cite{ERCP01}]
}
\end{table}

To heighten the contrast, the data in Fig.~\ref{lin-log} can be replotted (in Fig.~2b of Ref.~\cite{Hailu}) using as the ordinate $\varrho(\tilde{A}) \times \sigma^2(\tilde{A})$, so that approximations for which$\sigma^2 \propto \tilde{A}_{\rm eff}^{-1/2}$ become horizontal lines. 
With such rescaling, then, the Saclay and the
modified Grenoble (all steps interacting) predictions appear as lines at ordinates 0.405 and 0.495, respectively (cf.~Table 1).  The solid curve representing the GWD rises slowly, from 0.4 at $\tilde A$=1 toward 1/2, capturing a similar rise of the MC data.
\vspace{-5mm}
\section{User's guide: Synopsis of findings}
Refs.~\cite{Giesen01} and \cite{Giesen00} contain several figures showing applications, to experimental TWDs, of the perspective discussed above.  In the following we summarize several specific ideas that should be of use in confronting data.  

\noindent \textbf{Interaction strength from variance:}
From Eqs.\ (\ref{e:sr}) and (\ref{e:avrh}), one can estimate the variance from $\tilde{A}$, but
experimentalists usually seek the reverse, measuring the variance of the TWD and seeking to extract $A$. An excellent
estimate \cite{RCEG00} of the GWD-predicted
$\tilde{A}$ from the variance, based on series expansion of
Eq.\ (\ref{e:sr}), is:
\begin{equation}
  \label{e:as}
  \tilde{A}
    \approx  \frac{ 1}{16} \left[(\sigma^2)^{-2}
          - 7 (\sigma^2)^{-1}
          + \frac{27}{4}
          + \frac{35}{6} \sigma^2\right]  \, ,
\end{equation}
\noindent with {\it all four terms needed} to provide a good approximation
over the full physical range of $\tilde{A}$.  The Gaussian methods described
earlier essentially use just the first term of this expression and adjust the
prefactor.  

\noindent \textbf{Gaussian Fits of the GWD:}
Since TWDs for strong repulsions are well described by Gaussians, the GWD is well approximated by a Gaussian in
this limit.  In Ref.\ \cite{RCEG00} a quantitative assessment is given of how
closely the two distributions correspond as a function of $\varrho$.
At the calibration point (for which an exact
solution exists) for repulsive
interactions  ($\varrho =4$),  the relative difference of the standard deviation
of a Gaussian fitted to $P_{\varrho}(s)$ from the actual standard deviation of
this GWD (viz. the square root of the second moment of $P_{\varrho}(s)$ about its
mean of unity) is  $\sim$1\%,
and decreases monotonically with increasing
$\varrho$. (The $\chi^2$, however, is poorer with a Gaussian than with the GWD.)  For this range ($\varrho \geq 4$) differences between estimates of
$\tilde{A}$ obtained from GWD and the various Gaussian-fit
methods come primarily from different philosophies of
extracting $\tilde{A}$ from
$\sigma$ rather than from differences in the fitting schemes.

Gaussian approximations assume the peak (mode) is at $s = 1$ ($\ell = \la \ell \ra$); in fact, since  peak of the GWD must 
lie below one to achieve a mean of one; the tabulations in Table 1 bear this out.  Only to the extent that the mode is close to unity is a Gaussian approximation reasonable.  Formulas have been
derived \cite{RCEG00} indicating the errors in fitting $\varrho$ due to
errors in the first or zeroth moment of the distribution.
Another criterion for the validity of the Gaussian fitting function is that the TWD not be noticeably asymmetric about its peak, i.e.\ that it be negligible for $\ell > 2\la \ell \ra$. 

\noindent \textbf{Analyzing TWD skewness: Unfulfilled hopes:}
When $\varrho$ is too small for the TWD to meet the criteria for adequate fitting by a Gaussian, we hoped that one could obtain reliable estimates of $\tilde A$ by analyzing the skewness.  Although we tried a variety of formulas and schemes \cite{EP99,Giesen00}, it turned out that in the end a fit to the GWD was needed, so that skewness did not offer a shortcut to $\varrho$.

\noindent \textbf{Correcting estimates based on Gruber-Mullins:}
When dealing with tabulations of data analyzed in the traditional way
\cite{JW99} based on the Gruber-Mullins perspective, it is
useful to recast Eq.~(\ref{e:as}) in a form that indicates the factor by which
the estimate $\tilde{A}_W$ based on GWD exceeds the traditional estimate
$\tilde{A}_G$:
\begin{equation}
  \label{e:aa}
  \tilde{A}_W/\tilde{A}_G \: [\equiv A_W/A_G] \:
    \approx  3  -  21\sigma^2 +\frac{81}{4}\sigma^4 +\frac{35}{2}\sigma^6\, .
\end{equation}
\noindent Since $\tilde{A} =A\tilde \beta/(k_BT)^2$, the ratio of the physical interaction strengths is the same as that of the
dimensionless strengths.

\noindent \textbf{Alternative: fitting with Gamma distribution:}
Although the GWD is  a simple, single-parameter function, it is not a ``canned" distribution and that one needs to input gamma functions.  For the smaller values of $\varrho$ when a Gaussian is inappropriate, there is a preprogrammed function that is available: the Gamma distribution 
\begin{equation}
  \label{e:Gam}
  P_\Gamma(x) = \frac{x^{\alpha-1}{\rm e}^{-x/\theta}}{\Gamma(\alpha)\theta^\alpha}
\end{equation}
is serviceable if one recasts the data in terms of $x$ as $s^2$; one the can identify $\theta$ as $b_\varrho^{-1}$ and, more importantly, $\alpha$ as $(\varrho+1)/2$.  This approach has not yet been tested with actual data.

\noindent \textbf{Wigner Distribution as a 2-parameter fit:}
The GWDs giving the best fits of experimental TWDs sometimes
have  first moments that differ somewhat from the first moments 
of the data, especially in cases termed ``poor data" \cite{Giesen00,RCEG00}
which exhibit a small ``hump" at large values of $s$, beyond the mode.
Moreover, it may be desirable to 
determine the scaling length (the ``effective mean," which 
equals the first moment for ideal GWDs) and the
variance in a single fitting procedure rather than to predetermine this length
from the first moment.  
This ``refined" scaling implies that the argument of $P_{\varrho}$ should be
$\ell/\bar{\ell}$, where $\bar{\ell}$ denotes the characteristic length
determined  along with $\varrho$ in a two-parameter least-squares fit of the data
to a GWD, leading to the replacement \cite{note2param}:  
\begin{equation}
P_{\varrho}(s) \rightarrow (\langle \ell \rangle/\bar{\ell}) P_{\varrho}(s
\langle \ell \rangle/\bar{\ell}) \quad\mbox{i.e.}\quad 
(\langle \ell\rangle/\bar{\ell}) P_{\varrho}(\ell/\bar{\ell}).
\label{e:Psa}
\end{equation}
\noindent In the specific applications to data for copper \cite{ERCP01}, $\langle \ell \rangle/\bar{\ell}$ tends to be
greater than unity, typically by several percent, but it is unclear whether this
is true for semiconductors (or even for other metals).
In our Monte Carlo simulations \cite{Hailu}, where we have greater
control of purity and uniformity than in experiments, the optimal $\bar{\ell}$ is
essentially identical to $\langle \ell \rangle$: a two-parameter fit is unnecessary.

\noindent \textbf{Effects of Lattice Discreteness:}
For large values of $\tilde A$, the continuum picture breaks down and one must confront the discreteness of the lattice both in actual physical systems and in Monte Carlo simulations.  This problem is discussed extensively in Refs.\ \cite{ERCP01} and \cite{Hailu} but are not repeated here since the relevant values of $\tilde A$ are larger than found in many physical systems.  By adapting the celebrated VGL model \cite{VGL}, we have found \cite{Hailu} that the roughening criterion translates to 
$
\tilde A_R = \la \ell \ra^4/6. 
$
Note that vicinal surfaces are {\it rough}, in contrast to the flat terrace orientation.  Thus, for $\la \ell \ra$=3, e.g., when $\tilde A_R$ exceeds $\sim \! \! 14$ (so above the physical range), the vicinal orientation becomes a facet.  In some cases such as Si(113), the steps are exceedingly straight and uniformly spaced, making them excellent templates for growth of nanowires \cite{Himpsel}.  It is therefore likely that this surface is a facet rather than a rough vicinal. 

\noindent \textbf{Estimate of Number of Independent Measurements:}
In order to estimate uncertainties in the determination of the TWD and, ultimately,
$\tilde{A}$, it is important to have a realistic value of the number of {\it
independent} measurements, a number generally 
much smaller than the total number of measurements.  This problem is discussed in Ref.\ \cite{RCEG00}
The upshot is that the number of ``independent" terrace widths is reduced from $L_yN$ (the number of atomic spacings across the sampled area along the mean step direction times the number of steps) by up to nearly two orders of magnitude, due to slow decay of correlations perpendicular to the steps.  Thus, several STM images are needed to obtain decent statistics.
\vspace{-5mm}
\section{New developments and directions}

\noindent \textbf{Pair correlation function}

The TWD amounts to a many-particle correlation function since one insists that there be no step between 0 and $\ell$ (or $s$).  Instead one can study the step-step correlation function $h_\varrho(s)$, essentially the probability of finding a pair of steps separated by $\ell$ regardless of how many steps lie in between.  Until $\ell$ reaches $\sim \la \ell \ra$ there is little difference between the two, but then the pair correlation begins to rise and peak near $2\la \ell \ra$ and at subsequent multiples of $\la \ell \ra$, with a decreasing envelope around the oscillations, so that eventually $h \to \la \ell \ra^{-1}$.  For fitting experimental data (in this case, Si(111) at 1100$^\circ$C), we \cite{Saul} used an asymptotic expression \cite{GK01}
\begin{eqnarray}
\! \! \! \la \ell \ra h_\varrho(s) &\sim& -\frac{1}{\pi^2\varrho \,s^2}
+2\sum_{j=1}^\infty\frac{d_j^2(\varrho)}{(2\pi s)^{4j^2/\varrho}}\cos(2\pi j s),  \\
d_j(\varrho) &=& \Gamma\left(1+\frac{2j}{\varrho}\right)\prod_{m=1}^{j-1}\left(\frac{2 m}{\pi \varrho}\right) \Gamma^2\left(\frac{2\pi m}{\varrho}\right)\left(\frac{2 m}{\varrho}\right) \nonumber \end{eqnarray}
\noindent that works well for $s>1/2$, better than the conventionally used harmonic lattice approximation, since the conventional ``harmonic" approximation \cite{SB99} is insufficiently accurate.  For smaller $s$ we patched onto $a_\varrho s^\varrho$.  While Ha \cite{Ha95} derived a general exact solution for $h_\varrho(s)$, it is computationally intractable. 

\noindent \textbf{Fokker-Planck}

Recently, Pimpinelli et al. \cite{AP-FP}, starting from Dyson's 1D Coulomb gas model and making plausible assumptions, derived a Fokker-Planck equation
\begin{equation}
\! \! {\partial P(s,\tilde t)\over\partial \tilde t}={\partial \over\partial s}\left[\left({2b_\varrho}s-{\varrho\over s}\right)P(s,\tilde t)
\right]+{}{\partial^2 \over\partial s^2}[P(s,\tilde t)],
\label{e:FP}
\end{equation}
\noindent where $\tilde t \equiv t/\tau$, and the characteristic time $\tau$ is $\la \ell \ra^2$ over the squared strength of the noise in the underlying Langevin equation.  The steady-state solution of Eq.~(\ref{e:FP}) is the GWD.  With it we can describe with compact analytic expressions the
evolution toward equilibrium of steps from several experimentally relevant initial distributions: perfect cleaved crystals ($P(s) = \delta(s-1)$ as in Fig.~\ref{f:hiTWD}), step bunches, and prequench equilibrated distributions at different temperatures ($P_{\varrho_0}(s)$).  The decay time $\tau$ of the difference of variance of $P(s,t)$ from its saturation value, we have also found \cite{Ajmi}, can be related to underlying atomistic processes in kinetic Monte Carlo simulations of the evolution of an initially constrained vicinal surface.

These ideas have broad ramifications. In econophysics the same formalism arises for the distribution of the means of stock prices in the Heston model \cite{DY02}.  Similar behavior may occur in precipation patterns at geothermal hot springs \cite{GCV}.

The argument \cite{AP-FP} for Eq.(\ref{e:FP}) bolsters the formal justification for the GWD, as does work by Richards et al. \cite{beyond} that uses a
two-particle  Calogero model\cite{C69} (har- monically bound interacting spinless fermions
on a line). 

\noindent \textbf{Azimuthal Misorientation}

Most of the above has tacitly assumed that the steps are close-packed or at least oriented along high symmetry directions.  For vicinal surfaces surfaces that are misoriented in the azimuthal in addition to the polar orientation, there are complications in applying the GWD formalism, particularly in determining the dependence of $\tilde A$---ultimately of $\tilde \beta$ and $A$---on in-plane misorientation $\theta$.  For $\tilde \beta$ we have made substantial progress recently, again prompted by collaboration with the J\"ulich group \cite{Dielu} and by persistent probing questions by Ibach.  For the \{100\} and \{111\} faces of fcc cubic crystals, we have derived surprisingly simple formulas for the $\beta(\theta)$ and $\tilde \beta(\theta)$ for low $T$ (compared to the terrace roughening temperature) \cite{TJS100,TJS111}; in the case of Cu and other noble metals, this criterion is easily satisfied for room $T$.  For arbitrary $T$ we have generated a more complicated analytic expression that is straightforward incorporate in continuum simulations, in particular finite-element codes \cite{TJSsiam}.  We have also carried out ab initio calculations of the characteristic energies of the lattice-gas models used to understand $\tilde \beta(\theta)$ for Cu, to insure that these numbers are consistent with those estimated from experiments using statistical-mechanics reasoning \cite{TJSVASP}.  Fuller discussion is beyond the limits of the article.

Much less is known about $A(\theta)$, but again the J\"ulich group is at the forefront of these investigations \cite{GD}.  Another important open is the relation between $A$ and surface stress \cite{I97}.  There has been no successful extension of the seminal theory for isotropic substrates \cite{Noz,Mar} to account quantitatively for $A$ of a realistic surface \cite{KPK}.

\vspace{-5mm}
\section{Conclusion}
\label{sec-conclusion}

The GWD has proved a powerful tool to link the study of the step spacings on vicinal surfaces to very general ideas of fluctuations.  It is remarkable that H. Ibach could deduce its form, for the $A \! = \! 0$ case, from physical insight.  His perspicacious graph (Fig.~\ref{f:WI} spawned great progress in understanding and exploiting the fluctuations of the spacings between steps.  Even if one chooses to fit with a Gaussian, the theory of GWD provides the most reliable way to extract the strength of the step repulsions from the TWD.  In addition to clarifying the equilibrium properties of steps, these ideas are now helping us to understand non-equilibrium aspects, notably the relaxation of steps toward equilibrium.  Pimpinelli and I are also actively investigating applications to various aspects of growth in surface and interface problems.
\vspace{-5mm}
\section*{Acknowledgment}

My work has been supported primarily by the NSF-MRSEC at Univ.\ of Maryland (NSF Grants DMR 00-80008 and DMR 05-20471).  Collaborations with theorists N.C.\ Bartelt, B.~Jo\'os, O.\ Pierre-Louis, H.L.\ Richards, A.~Pimpinelli have been indispensible, as has the assistance of graduate students Hailu Gebremariam and T.J. Stasevich, and undergraduates S.D.\ Cohen and R.D.\ Schroll.  Collaboration with H. Ibach and M. Giesen was fostered by an Alexander von Humboldt Foundation U.S. Senior Scientist Award and the kind hospitality of IGV and then ISG during several visits at the Forschungszentrum J\"ulich.
At Maryland I have long enjoyed collaborations with the large surface experimental group, headed by Ellen~D.\ Williams, and also with J.-J. M\'etois in Marseilles.  I have also benefited from interactions on the GWD with many others, particularly often with H.\ van Beijeren and M.E. Fisher.  
\vspace{-5mm}
\section*{Appendix: How the Hamiltonian symmetries lead to the Wigner surmise exponents}
This appendix, based on Ref.~\cite{Zwer98} (see also \S 4.3 of \cite{Haake91}), describes how the exponents of $s$ in Eqs.~(\ref{e:w2}--\ref{e:w4}) come about.
For physicists, random matrix theory is rooted in the study of energy eigenvalues in nuclear physics.  As recounted beautifully in Ref.~\cite{Guhr98}, conventional statistical mechanics treats ensembles of identical physical systems with the same Hamiltonian but different initial conditions.  In contrast, Wigner considered ensembles of dynamical systems governed by different Hamiltonians ${\cal H}$ having some common symmetry property and sought generic properties of such ensembles associated with the symmetry. Dyson extended this work to show that there are three symmetries of the matrices of the Hamiltonian:  orthogonal (real symmetric) matrices correspond to time-reversal invariance with rotational symmetry, unitary (complex) matrices to violated time-reversal invariance (as for electrons in a magnetic field), symplectic (see Eq.~(2.3) of \cite{Guhr98}, \S 2.4 of \cite{MehtaRanMat}) to time-reversal invariance with broken rotational symmetry and 1/2-integer spin.
Wigner then, for convenience, considered Gaussian weightings, 
\begin{equation}
p({\cal H}) \sim \exp [-b N\, {\rm tr}({\cal H}^2)],
\label{e:WD}
\end{equation}
\noindent with the idea that for large matrices the {\it fluctuations} of the eigenvalues should be independent of the weight factors as well as of the specific form of the level spectrum.  (There is no information about the {\it average} values.)  

As described in Ref.~\cite{Zwer98}, we start with the simplest case, an orthogonal ensemble ($\beta$=1)
with $N$=2 (just 2 particles), with ${\cal H}$ having diagonal elements $h_{11}$ and $h_{22}$, and off-diagonal element $h_{12} (=h_{21})$.
We focus on the Jacobean
associated with a change of variables for the Gaussian-ensemble
probability distribution function in going from the joint probability distribution function $p(E_1,E_2)$ for adjacent eigenenergies $E_1,E_2$ to $P(s) ds$; we define variables $\bar{h} \equiv \left( h_{11}+ h_{22}\right)/2$, $u\equiv h_{11}-h_{22}$ and $s\equiv\left( u^2 + 4h_{12}^2\right)^{1/2} =|E_2-E_1|$.  Thus, $E_{1,2} = \bar{h} \pm s/2$.   We must now take into account all possible matrix elements $h_{11}, h_{22}, h_{12}$. From Eq.~(\ref{e:WD}) and with $dh_{11}dh_{22} = d\bar{h} du$, 
\begin{equation}
\int \! \! \! \! \! \int \!\! \! \! \!  \int\! \!  
p\, dh_{11}  dh_{22}  dh_{12} \! = \! \int \! \! \! ds \! \!  \int \!  \! \! \! \! \int \! \exp\! \left[\! -2b(E_1^2\! +\! E_2^2)\right] \! d\bar{h} du \! \left| \frac{dh_{12}}{ds}\right|
\label{e:Whs}
\end{equation}

Hence we can identify the inner double integral of Eq.~(\ref{e:Whs}) as $p(s)$.  
Since $h_{12} = \pm(1/2)(s^2-u^2)^{1/2}$, $\left| dh_{12}/ds\right| = (s/2)\left(s^2-u^2\right)^{-1/2}$.
Since $E_{1,2}$ do not depend on $u$, we can pull the exponential out of the integration over $u$, leaving the elementary integral

\begin{equation}
(s/2)\int_{-s}^{s} \left(s^2-u^2\right)^{-1/2} du = \pi s/2 
\label{e:Wu}
\end{equation}
Next, since $2(E_1^2+E_2^2) = s^2 + 4\bar{h}^2$, the integration over $\bar{h}$ is also elementary, and we are left with 
$P(s) \sim s \exp(-bs^2)$.  This exact result for $N=2$ provides an excellent approximation for large $N$ as well.  Indeed, near the level crossing (corresponding to $s \rightarrow 0$) the problem tends to reduce to a $2\times 2$ problem.
The integral for $P(s)$ over variables $u$ and $h_{12}$ includes in its integrand a Dirac delta function $\delta\left( s -[u^2 +4h_{12}^2]^{1/2}\right)$, which
vanishes for $s$=0 only when the two [squared] independent
variables do.  Hence, $P(s) \propto s$, corresponding to a
circular shell in parameter space, and we again have $\beta$=1.  

For
unitary ensembles there is an additional independent parameter
since $h_{12}^2$ becomes $(\Re {\rm e}\, h_{12})^2 + (\Im {\rm
m}\, h_{12})^2$. Hence, $P(s) \propto s^2$, corresponding to a
spherical shell in parameter space, i.e.\ $\beta$=2. The argument for the 
symplectic ensemble leads similarly---but via quaternions or Pauli spin matrices---to $\beta$=4.
\vspace{-6mm}


\begin{thebibliography}{10}
\bibitem{MC50} For $\tilde{A}\! = \! 50$, we found it safest to increase to $L_y \! = \! 3000$,  $N \! = \! 200$.
\bibitem{note2param} Since $s$ is still determined from the raw data as 
$\ell/\langle \ell \rangle$, the refined scaling translates into replacing
$s$ by $s \langle \ell \rangle/\bar{\ell}$ in the argument of the distribution. 
If the integration variable $s$ were similarly replaced, then the refined
scaling would amount to a redefinition of a dummy variable, normalization
would still be realized; however, the independent variable is kept as
$s$ \cite{ERCP01}.
\end{thebibliography}
\end{document}